\documentclass[11pt]{article}

\usepackage{arxiv}

\usepackage[utf8]{inputenc}
\usepackage[T1]{fontenc}
\usepackage{hyperref}
\usepackage{url}
\usepackage{booktabs}
\usepackage{amsfonts}
\usepackage{amsmath}
\usepackage{amssymb}
\usepackage{nicefrac}
\usepackage{microtype}
\usepackage[ruled,vlined,linesnumbered]{algorithm2e}
\SetAlgoLined
\usepackage{pifont}
\newcommand{\cmark}{\ding{51}}
\newcommand{\xmark}{\ding{55}}
\usepackage{graphicx}
\usepackage{xcolor}
\usepackage[numbers,sort&compress]{natbib}

\hypersetup{
  colorlinks=true,
  linkcolor=blue!70!black,
  citecolor=blue!70!black,
  urlcolor=blue!70!black
}

\title{ScoreGate: Adaptive Chunk Selection for\\
       Retrieval-Augmented Generation\\
       via Dual-Score Statistical Fusion}

\author{
  \begin{tabular}[t]{@{}l@{\hspace{4em}}l@{}}
    Karamvir Singh\thanks{Correspondence: \texttt{karamvir.singh@gohighlevel.com}} &
    Arvind Jain\thanks{\texttt{arvind.jain@gohighlevel.com}} \\[0.2em]
    \small\itshape Staff AI Engineer &
    \small\itshape Senior Engineering Manager \\[0.1em]
    \small HighLevel, Inc. &
    \small HighLevel, Inc.
  \end{tabular}
}

\begin{document}
\maketitle

\begin{abstract}
Fixed-cardinality retrieval injects a constant top-$K$ chunks into the
generator regardless of query complexity, causing over-retrieval for
narrow queries and under-retrieval for compositional ones.
We describe \textbf{ScoreGate}, a lightweight score-space decision
mechanism that controls \emph{retrieval cardinality} at inference time
using two scores already produced by the standard pipeline: bi-encoder
similarity $s_i$ and cross-encoder reranker score $r_i$, with no
additional model inference calls required.
Its core insight is that cross-encoder affirmation can rescue
semantically relevant chunks that bi-encoder retrieval ranks poorly
due to vocabulary mismatch---a failure mode unaddressed by fixed-$K$
or single-score thresholding.
On MS~MARCO (200 dev queries), ScoreGate achieves MRR@10\,=\,0.401
with 35\% fewer retained chunks than Standard Top-K.
On an internal benchmark ($n\!=\!300$, Fleiss' $\hat{\kappa}\!=\!0.87$),
ScoreGate observed zero false positives (95\% CI [96.4\%, 100\%]) at
97.77--99.34\% recall, with 34.8\% fewer tokens per query and only
31\,ms added latency.
Results on both MS~MARCO and real-world production traffic suggest
that adaptive retrieval cardinality can improve retrieval efficiency
without degrading retrieval quality.
\end{abstract}


\section{Introduction}

Retrieval-Augmented Generation (RAG) augments a language model prompt with
passages retrieved from an external corpus, enabling document-grounded
responses~\citep{Lewis2020}.
The retrieval stage operates in two phases: a \emph{bi-encoder} embeds the
query and corpus documents into a shared vector space and retrieves the
top-$N$ candidates by cosine similarity; a \emph{cross-encoder reranker}
scores each candidate by performing full query--document attention, producing
measurably more accurate relevance estimates than embedding similarity
alone~\citep{Nogueira2019,Khattab2020}.
The generator then receives a fixed-cardinality subset of top-$K$ chunks,
where $K$ is a hyperparameter set before deployment.

The fixed-$K$ constraint imposes a hard upper bound on retrieval cardinality
that is mismatched to query complexity.
A navigational query such as ``\emph{What is the refund policy?}''
is fully answered by a single chunk, whereas a compositional query such as
``\emph{Compare plans A and B across pricing, features, and cancellation
terms}'' may require eight or more chunks.
Setting $K$ conservatively causes coverage gaps on broad queries;
setting it liberally injects low-relevance distractors that increase
generation error rate and token cost~\citep{Wang2023,Wang2023b}.
Neither a larger nor a smaller $K$ resolves this tension because the optimal
cardinality is a function of the query, not a constant.

Existing approaches address retrieval \emph{quality} but
generally do not treat cardinality as a first-class decision.
Rerankers reorder candidates but do not select a subset.
LLM-based chunk filters~\citep{Wang2023b} prompt a language model for a binary
keep/discard decision per chunk, achieving good precision but introducing an
additional inference call per query.
Single-score threshold cutoffs are unreliable: high embedding similarity does
not guarantee contextual relevance, and low embedding similarity does not
exclude relevance (paraphrastic references use different vocabulary).

We observe that the \emph{joint} score pair $(s_i, r_i)$ carries
measurably richer signal than either score in isolation.
The two scores are complementary: $s_i$ captures lexical and semantic
proximity in the embedding space, while $r_i$ captures contextual fit via
full cross-attention.
Chunks where both scores agree can be classified deterministically;
chunks where they disagree require an empirically motivated fusion strategy.
We observe a further structural failure in standard pipelines:
bi-encoder retrieval systematically under-ranks chunks that are
semantically relevant but expressed using vocabulary absent from
the query (paraphrases, domain-specific synonyms, indirect
references).
The cross-encoder reranker, which performs full query--document
attention, often correctly scores these chunks as relevant despite
their low embedding similarity.
A single-score threshold on either $s_i$ or $r_i$ cannot recover
these chunks reliably.
The joint score pair $(s_i, r_i)$ carries the signal needed to
identify and rescue them.
This motivates \textbf{ScoreGate}.

\paragraph{Design and contributions.}
ScoreGate is a lightweight score-space decision mechanism rather
than a new retrieval architecture.
It slots into any existing two-stage RAG pipeline without
architectural changes, additional training, or extra model
inference calls.
Its main design contributions are:
\begin{itemize}
  \item \textbf{Dual-score bucket classification.}
        A four-region partition of the $(s_i, r_i)$ score space, with
        deterministic rules for agreement regions and threshold-based fusion
        for disagreement regions---requiring no model calls beyond the
        standard reranker.

  \item \textbf{Asymmetric fusion thresholds.}
        Bucket-specific retention thresholds
        ($\theta_{B2}\!=\!0.255$, $\theta_{B3}\!=\!0.15$) derived from the
        observation that reranker-affirmed chunks in B3 are more likely true
        positives than reranker-rejected chunks in B2 at equivalent fusion
        scores.

  \item \textbf{Empirical validation on real-world data.}
        Observed precision within 95\% CI [96.4\%, 100\%] at 97.77--99.34\% recall (vs.\ 91.11--92.41\% for LLM
        filtering), 34.8\% token reduction, and only 31\,ms added latency on
        a 300-triple annotated benchmark and a real-world query dataset of
        1,247 live queries.

  \item \textbf{Production safety.}
        A MAX-$K$ ceiling prevents context-window overflow and preserves
        backward compatibility with fixed-context downstream components.
\end{itemize}

\section{Related Work}

\subsection{Retrieval-Augmented Generation}
Lewis et al.~\citep{Lewis2020} established RAG as a paradigm for knowledge-intensive NLP.
Subsequent work extended RAG to iterative retrieval--generation
loops~\citep{Trivedi2022,Shao2023} and pre-/post-retrieval optimisation
strategies~\citep{Lin2023}.
Wang et al.~\citep{Wang2023} showed that injecting low-relevance passages degrades
generation quality proportionally to the fraction of irrelevant content in
the context window; they address this via learned context filtering at generation
time.
A concurrent line of work~\citep{Wang2023b} addresses the same problem via
query-oriented passage scoring before the generator sees any context.
ScoreGate is complementary to both: we determine \emph{how many} chunks
to pass to the generator, rather than filtering or reranking within a
fixed-cardinality set.

\subsection{Bi-Encoder Retrieval and Cross-Encoder Reranking}
Bi-encoders retrieve by nearest-neighbor search over pre-computed document
vectors~\citep{Nogueira2019}.
Cross-encoder rerankers compute a relevance score via full self-attention
over the query--document pair, producing more accurate estimates at higher
cost~\citep{Nogueira2019}.
ColBERT~\citep{Khattab2020} achieves a favourable accuracy--efficiency
trade-off using token-level late interaction.
In production RAG, the standard architecture is bi-encoder retrieval of
top-$N$ candidates followed by cross-encoder reranking to produce
top-$K$ chunks for the generator.
ScoreGate intervenes at this second stage, replacing fixed-$K$ truncation
with score-adaptive selection.

\subsection{Adaptive and Conditional Retrieval}
FLARE~\citep{Jiang2023} conditions retrieval triggering on generator
token-level confidence.
Self-RAG~\citep{Asai2024} trains the generator to emit control tokens
indicating retrieval necessity and passage relevance.
Both methods address retrieval \emph{triggering} (when to retrieve);
ScoreGate addresses retrieval \emph{cardinality} (how many to
retain)---orthogonal problems that can be composed.

\subsection{Adaptive Retrieval Cardinality}

A small body of work addresses dynamic stopping and cardinality
control in retrieval.
\emph{Score-gap stopping}~\citep{Wang2023} truncates the ranked
list when the score gap $r_i - r_{i+1}$ exceeds a threshold,
exploiting the intuition that large gaps signal relevance
boundaries.
\emph{Confidence-based retrieval} in Self-RAG~\citep{Asai2024}
conditions the number of retrieved passages on token-level
generation confidence.
\emph{Adaptive truncation} methods~\citep{Asai2024} determine
retrieval necessity from entity popularity, bypassing cardinality
decisions for well-memorised facts.
ScoreGate is complementary to these approaches: it operates
after reranking using scores already computed, requires no
additional model calls, and specifically targets the
vocabulary-mismatch recovery problem unaddressed by gap-based
or confidence-based methods.

Table~\ref{tab:comparison} positions ScoreGate relative to
common retrieval methods.

\begin{table}[h]
  \caption{
    \textbf{Method comparison.}
    ScoreGate is the only method that provides adaptive
    cardinality using a reranker without extra inference.
  }
  \label{tab:comparison}
  \centering
  \begin{tabular}{lcccc}
    \toprule
    \textbf{Method} & \textbf{Type} &
    \textbf{Reranker} & \textbf{Adaptive card.} &
    \textbf{Extra inference} \\
    \midrule
    Fixed Top-K    & Truncation & \xmark & \xmark & \xmark \\
    Score-gap stop & Truncation & \cmark & \cmark & \xmark \\
    LLM Filter     & Filter     & \cmark & \cmark & \cmark \\
    Self-RAG       & Gen.\ ctrl & \xmark & \cmark & \cmark \\
    \textbf{ScoreGate} & Filter & \cmark & \cmark & \xmark \\
    \bottomrule
  \end{tabular}
\end{table}

\subsection{Multi-Signal Score Fusion}
CombSUM, CombMNZ~\citep{Fox1994}, and Reciprocal Rank
Fusion~\citep{Cormack2009} combine ranked lists from multiple retrieval
systems for rank aggregation.
ScoreGate instead fuses two continuous scores from a single candidate set
to make a \emph{binary} keep/discard decision---a fundamentally different
application of score fusion.

\section{Problem Formulation}

Let $Q$ denote a user query.
A two-stage RAG pipeline produces a candidate set
$C = \{c_1, \ldots, c_N\}$ of the top-$N$ chunks ranked by the
cross-encoder reranker.
Each chunk $c_i$ is associated with a score pair:
\begin{equation}
  c_i = (s_i,\ r_i)
  \label{eq:pair}
\end{equation}
where $s_i \in [0,1]$ is the bi-encoder cosine similarity between the
$L_2$-normalised query and chunk embeddings, and $r_i \in [0,1]$ is the
cross-encoder relevance score normalised to $[0,1]$ by min--max scaling
over $C$.
Per-query min--max normalisation is intentional: the thresholds
$\tau_s$ and $\tau_r$ are derived from the same normalised distributions,
so they are internally consistent.
More importantly, the relative ordering of chunks within a query is
preserved under any monotone transformation.
A chunk classified as B3 ($r_i \geq \tau_r$) is affirmed by the
cross-encoder \emph{relative to the other candidates for that query}---
the signal we want: whether the reranker considers this chunk relevant
given the competition, not whether it exceeds an absolute threshold.
The false-positive rate of 5\% used to derive $\tau_r$ (Section~4.1)
was computed on the same per-query normalised scores, so the 5\% FPR
claim is consistent with this normalisation.
Table~\ref{tab:norm} confirms empirically that per-query min--max
scaling outperforms raw logits and sigmoid normalisation on both
ARB recall and MS~MARCO MRR@10.

\begin{table}[h]
  \caption{
    \textbf{Reranker score normalisation ablation.}
    Per-query min--max scaling outperforms alternatives by
    producing the strongest B3/B4 score separation.
    Results on ARB (Sem.\ Relevant recall) and MS~MARCO
    (200-query dev set, MRR@10).
  }
  \label{tab:norm}
  \centering
  \begin{tabular}{lccc}
    \toprule
    \textbf{Normalisation} &
    \textbf{Recall (Sem.\ Rel.)} & \textbf{MRR@10} &
    \textbf{Avg.\ $|C'|$} \\
    \midrule
    Raw logits            & 96.8\% & 0.392 & 7.1 \\
    Sigmoid               & 97.5\% & 0.395 & 6.8 \\
    \textbf{Min--max (ours)} & \textbf{99.34\%} & \textbf{0.401} &
    \textbf{6.1} \\
    \bottomrule
  \end{tabular}
\end{table}

Per-query min--max normalisation produced the strongest separation
between B3 and B4 regions, likely because it preserves within-query
score ranking while adapting to query-specific reranker score
distributions.
The standard pipeline selects
$C' = \{c_{(1)},\ldots,c_{(K)}\}$ by fixed truncation at rank $K$.
We seek instead an adaptive selection function $F$ such that:
\begin{equation}
  F(C) \rightarrow C' \subseteq C,\quad 1 \leq |C'| \leq K_{\max}
  \label{eq:goal}
\end{equation}
where $F$ requires no model inference beyond the reranker call already
performed, and $|C'|$ is determined by the score distribution of $C$
rather than fixed a priori.

\section{ScoreGate Method}

\begin{figure}[t]
  \centering
  \includegraphics[width=\textwidth]{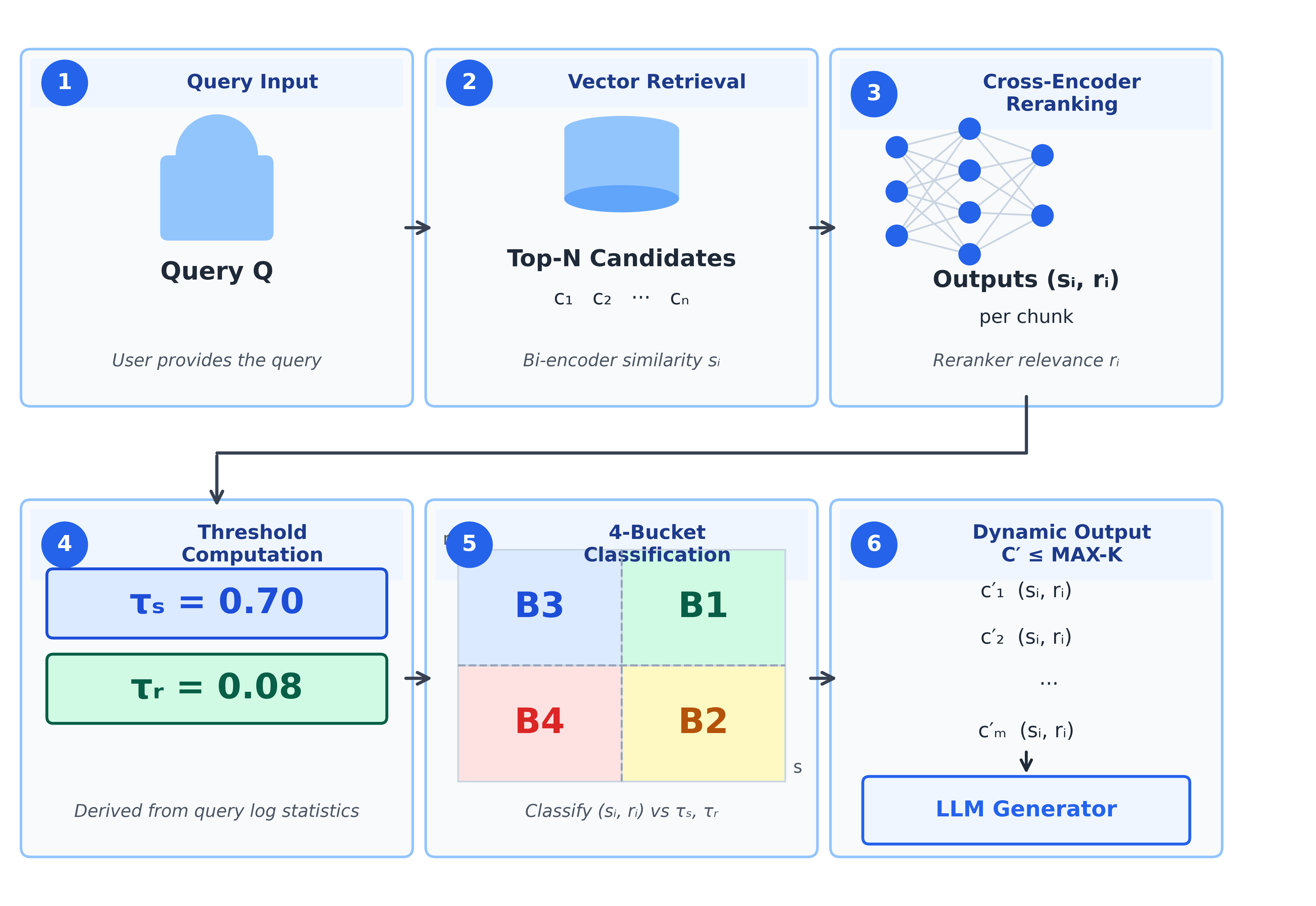}
  \caption{
    \textbf{The ScoreGate pipeline.}
    A user query triggers bi-encoder vector retrieval of top-$N$ candidates,
    followed by cross-encoder reranking to obtain score pairs $(s_i, r_i)$.
    Data-driven thresholds ($\tau_s\!=\!0.70$, $\tau_r\!=\!0.08$) partition
    each chunk into one of four buckets (B1--B4).
    B1 and B3 are retained unconditionally; B4 is discarded
    unconditionally; B2 candidates are retained only if the fusion score
    $f_i \geq 0.255$.
    The filtered set $C'$ is bounded above by MAX-$K$ and passed to
    the generator.
  }
  \label{fig:pipeline}
\end{figure}

Figure~\ref{fig:pipeline} illustrates the complete ScoreGate pipeline.
We use the term \emph{adaptive cardinality} rather than \emph{filtering}
to emphasise that ScoreGate does not re-rank or post-process a
fixed-size set---it determines the \emph{size} of the set passed to
the generator, which can be anywhere from 0 to MAX-$K$ depending on
the query's score distribution.
The distinction matters: a filter operates on a fixed-$K$ input;
ScoreGate replaces the fixed-$K$ decision itself.

\subsection{Score Threshold Derivation}

We obtain thresholds $\tau_s$ and $\tau_r$ from query log statistics.
$\tau_s$ is the median bi-encoder cosine similarity over the top-$N$
candidate set aggregated across a representative query sample ($N\!=\!40$).
$\tau_r$ is the reranker score corresponding to a false-positive rate of
approximately 5\% on a held-out relevance-annotated validation set.
From our production logs:
\begin{equation}
  \tau_s = 0.70 \qquad \tau_r = 0.08
  \label{eq:thresholds}
\end{equation}
Both thresholds are computed once from log data and applied at inference
time without per-query recalculation.
The threshold derivation procedure is domain-agnostic: re-running the
median-similarity and 5\%-FPR calibration on a domain-specific query
sample produces new $\tau_s$, $\tau_r$ values without any model
retraining. Whether the four-bucket structure and fusion weight
$\alpha\!=\!0.3$ generalise across domains is an empirical question
this is evaluated empirically in Section~4.3 (Figure~\ref{fig:alpha}).

\begin{figure}[h]
  \centering
  \includegraphics[width=\textwidth]{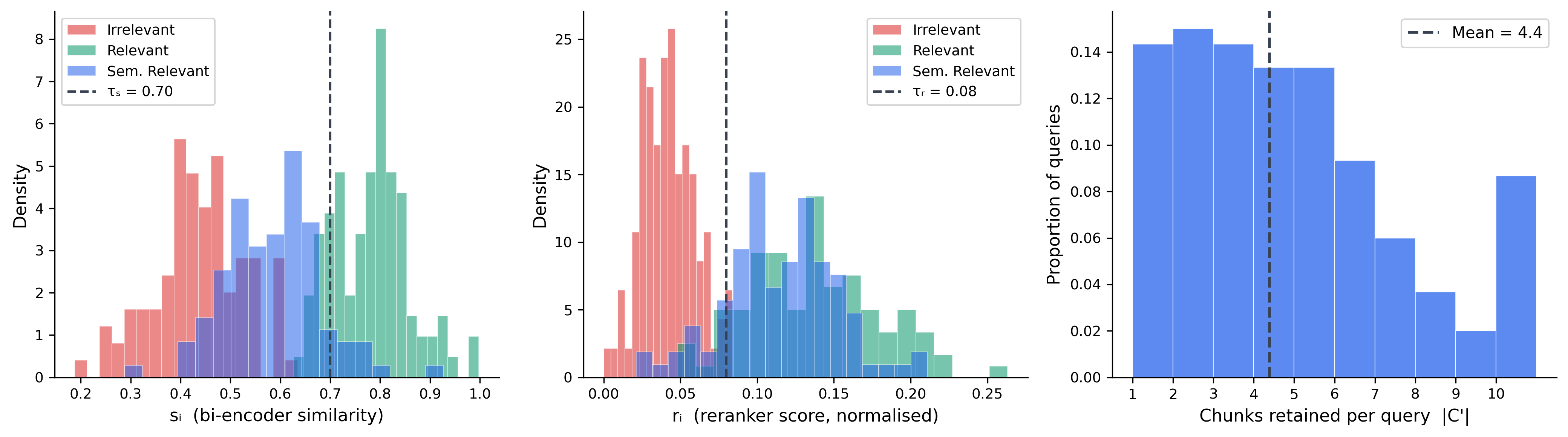}
  \caption{
    \textbf{Left/Centre:} Score distributions by relevance category
    on the ARB. Dashed lines mark $\tau_s\!=\!0.70$ and
    $\tau_r\!=\!0.08$. Semantically Relevant chunks (blue) have
    elevated $r_i$ despite low $s_i$---the B3 pattern.
    \textbf{Right:} Distribution of retained chunks per query
    ($|C'|$) on the ARB, demonstrating adaptive cardinality
    (mean 4.4 vs fixed $K\!=\!10$).
  }
  \label{fig:distributions}
\end{figure}

\paragraph{Score distributions by relevance category.}
Figure~\ref{fig:distributions} shows empirical score distributions
for $s_i$, $r_i$ across the three ARB relevance categories.
Irrelevant chunks cluster at low $s_i$ and low $r_i$; Relevant
chunks at high $s_i$ and high $r_i$.
Semantically Relevant chunks span a wide $s_i$ range but maintain
elevated $r_i$---confirming the B3 pattern that motivates
ScoreGate's asymmetric thresholds.
The right panel shows the distribution of retained chunk counts
$|C'|$ across ARB queries, confirming adaptive behaviour: most
queries retain 2--5 chunks rather than a fixed $K\!=\!10$.

\subsection{Four-Region Score Partition}

\begin{figure}[t]
  \centering
  \includegraphics[width=0.62\textwidth]{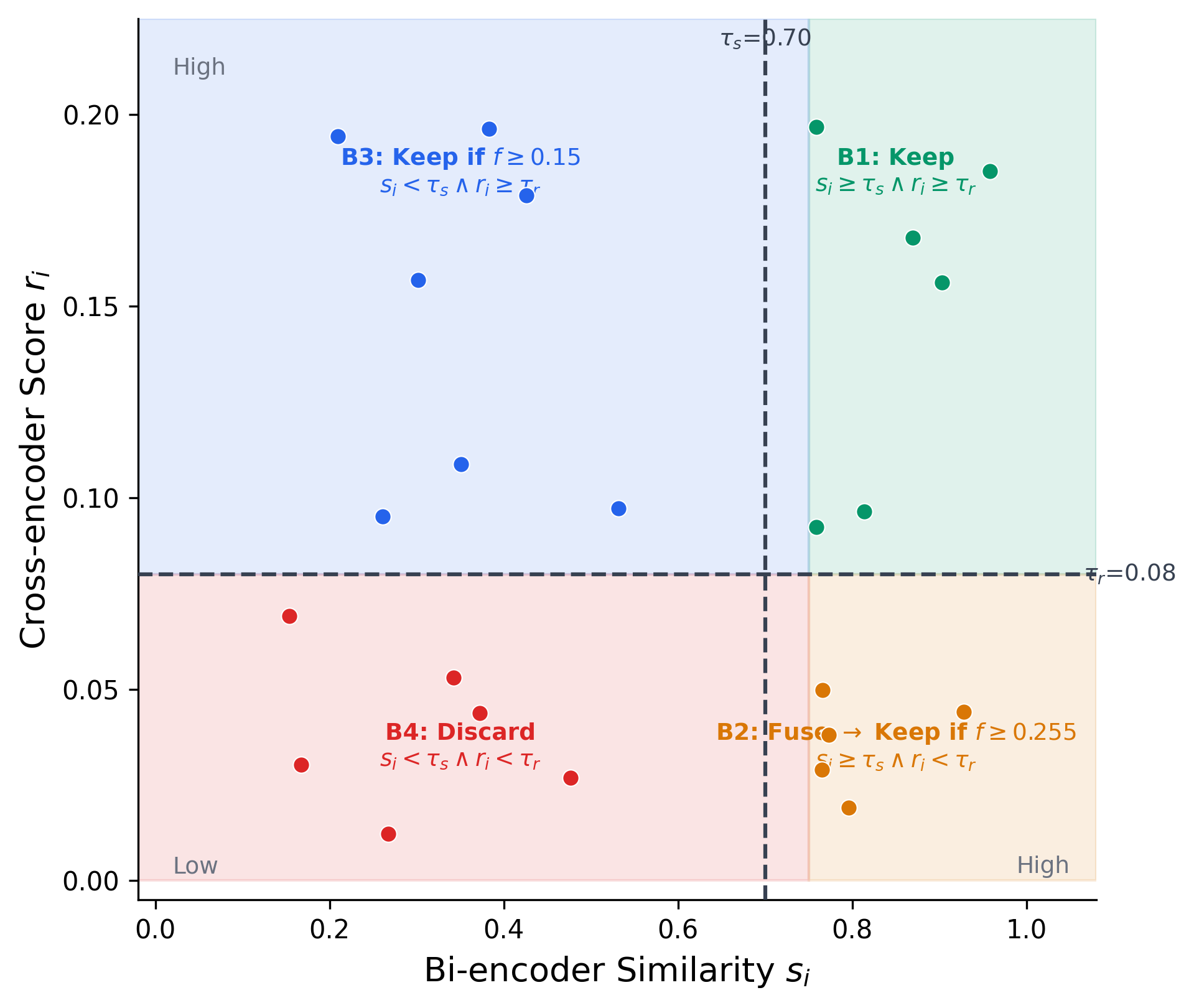}
  \caption{
    \textbf{Four-region partition of the $(s_i, r_i)$ score space.}
    Vertical dashed line at $\tau_s\!=\!0.70$;
    horizontal dashed line at $\tau_r\!=\!0.08$.
    \textbf{B1} (green, top-right): both scores affirm relevance---keep
    unconditionally.
    \textbf{B2} (orange, bottom-right): high similarity, low
    relevance---apply fusion threshold $f_i \geq 0.255$.
    \textbf{B3} (blue, top-left): low similarity, high relevance
    (vocabulary mismatch)---apply lower threshold $f_i \geq 0.15$.
    \textbf{B4} (red, bottom-left): both scores reject---discard
    unconditionally.
    Data points are representative examples from the Annotated
    Relevance Benchmark (ARB).
  }
  \label{fig:quadrant}
\end{figure}

The thresholds $\tau_s$ and $\tau_r$ partition the unit square
$[0,1]{\times}[0,1]$ of score pairs into four axis-aligned,
non-overlapping regions.
Each candidate $(s_i, r_i)$ falls into exactly one bucket, as shown in
Figure~\ref{fig:quadrant} and Table~\ref{tab:buckets}.

\begin{table}[h]
  \caption{
    \textbf{Four-region score-space partition and retention rules.}
    The B3 threshold is intentionally lower than B2 because the
    cross-encoder has already affirmed relevance in B3; the only
    disconfirming signal is lexical distance, which we treat as
    less informative than cross-encoder affirmation.
  }
  \label{tab:buckets}
  \centering
  \begin{tabular}{llll}
    \toprule
    \textbf{Region} &
    \textbf{$s_i$ condition} &
    \textbf{$r_i$ condition} &
    \textbf{Retention rule} \\
    \midrule
    B1 & $\geq\tau_s$ & $\geq\tau_r$ & Always keep \\
    B2 & $\geq\tau_s$ & $<\tau_r$    & Keep if $f_i \geq 0.255$ \\
    B3 & $<\tau_s$    & $\geq\tau_r$ & Keep if $f_i \geq 0.15$ \\
    B4 & $<\tau_s$    & $<\tau_r$    & Always discard \\
    \bottomrule
  \end{tabular}
\end{table}

\subsection{Weighted Fusion Score and Asymmetric Thresholds}

For candidates in B2 and B3, a deterministic rule based on a single score
is unreliable because the two scores carry conflicting evidence.
We compute a weighted linear fusion score:
\begin{equation}
  f_i = \alpha \cdot s_i + (1-\alpha) \cdot r_i, \qquad \alpha = 0.3
  \label{eq:fusion}
\end{equation}
The weight $\alpha\!=\!0.3$ assigns 70\% of the fusion weight to $r_i$
and 30\% to $s_i$.
This asymmetry is justified by two properties of cross-encoder rerankers:
(1) cross-encoders are trained directly on human relevance judgements via
supervised contrastive learning over query--document pairs, producing scores
better calibrated to human relevance than cosine similarity in the
bi-encoder embedding space;
(2) cross-encoder scores are conditioned on the specific query, whereas
bi-encoder similarity measures proximity in a shared space optimised for
retrieval recall rather than precision.

The retention thresholds $\theta_{B2}\!=\!0.255$ and
$\theta_{B3}\!=\!0.15$ are asymmetric by design.
In B2, the cross-encoder has rejected the chunk ($r_i < \tau_r$), so a
higher fusion score is required to override that rejection.
In B3, the cross-encoder has affirmed the chunk ($r_i \geq \tau_r$), and
the only disconfirming signal is low bi-encoder similarity---a ``lexical
gap'' we treat as less informative than cross-encoder affirmation.

\begin{figure}[h]
  \centering
  \includegraphics[width=0.82\textwidth]{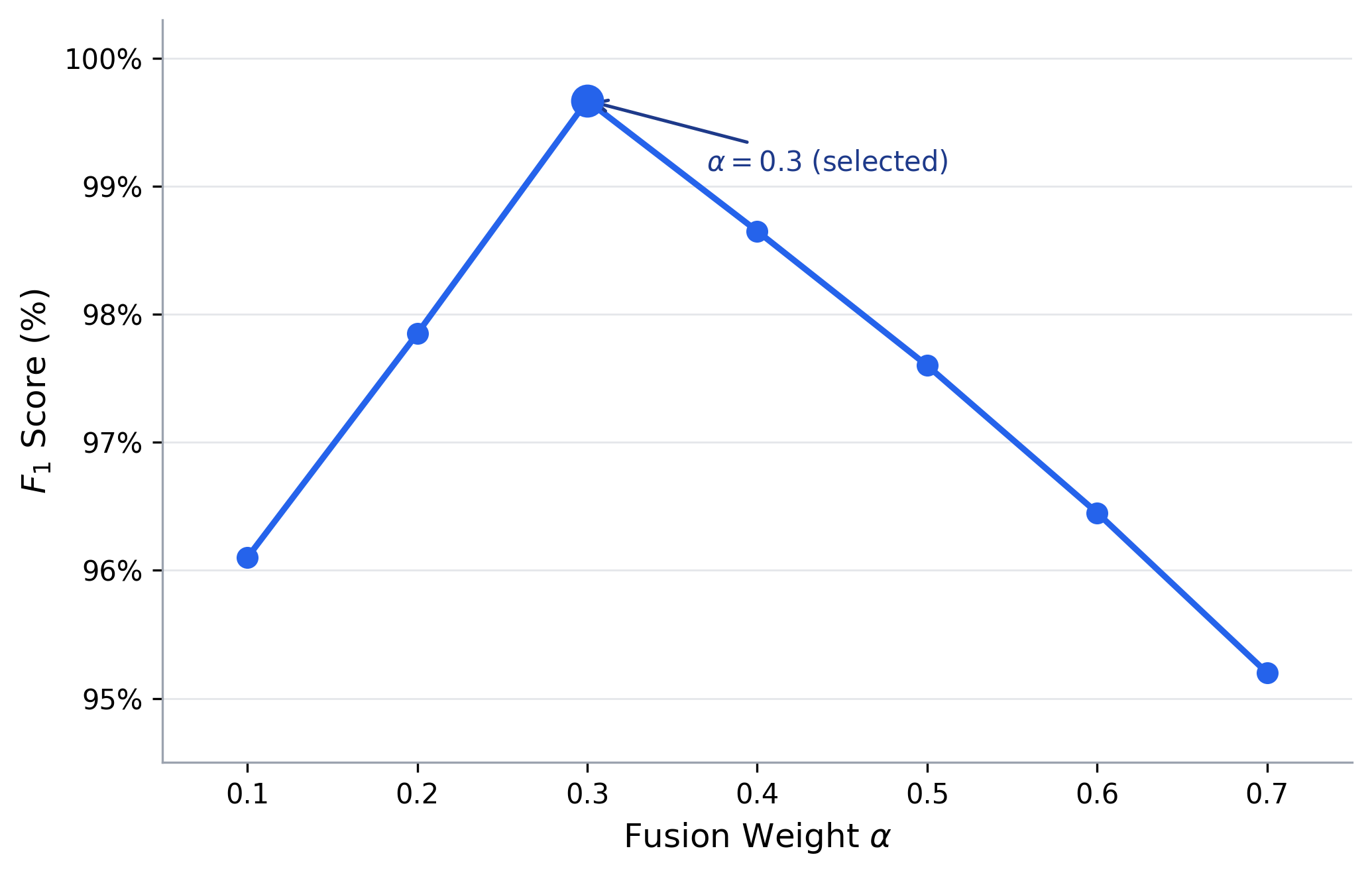}
  \caption{
    \textbf{Fusion weight sensitivity (ARB, Semantically Relevant).}
    $F_1$ peaks at $\alpha\!=\!0.3$ and declines monotonically in both
    directions, confirming that reranker-dominant weighting (70\% to
    $r_i$, 30\% to $s_i$) is empirically necessary.
    The unimodal shape indicates the optimum is stable under evaluated conditions: performance
    degrades gracefully rather than catastrophically away from
    $\alpha\!=\!0.3$.
  }
  \label{fig:alpha}
\end{figure}

Figure~\ref{fig:alpha} shows $F_1$ sensitivity to $\alpha$
across seven values; the surface peaks at $\alpha\!=\!0.3$ and
declines monotonically.

\subsection{Cardinality Bound}

After applying the four-region rules, the retained set $C'$ is bounded by:
\begin{equation}
  |C'| \leftarrow \min(|C'|,\ K_{\max})
  \label{eq:cap}
\end{equation}
When the ceiling is binding ($|C'| > K_{\max}$), we retain the $K_{\max}$
chunks with the highest fusion scores $f_i$.
This ranking is empirically motivated: $f_i$ is the same weighted combination
of both relevance signals used throughout ScoreGate, so the dropped
chunks are precisely those with the weakest combined evidence for
relevance among an already-filtered set.
In our deployment $K_{\max}\!=\!10$.
The ceiling was binding in 14 of the 300 ARB queries (4.7\%),
resulting in 31 dropped chunks.
Of these, 23 were Irrelevant (correctly dropped), 4 were Relevant,
and 4 were Semantically Relevant (8 total false negatives introduced
by the ceiling).
By bucket, dropped chunks were: B2\,=\,4, B3\,=\,9, B4\,=\,18
(no B1 chunks were dropped).
The 9 dropped B3 chunks are the most operationally significant:
these are vocabulary-mismatch chunks affirmed by the reranker but
with lower $f_i$ than other retained chunks due to their low $s_i$.
This confirms the vocabulary-mismatch edge case described in Section~4.1 and motivates
raising MAX-$K$ in knowledge-intensive deployment contexts.

\subsection{Algorithm}

The complete ScoreGate procedure is summarised in
Algorithm~1.
Time complexity is $O(N \log N)$ dominated by the final sort.
With $N\!=\!40$, this contributes negligible overhead relative to
cross-encoder inference.

\begin{algorithm}[h]
\DontPrintSemicolon
\caption{ScoreGate}
\label{alg:scoregate}
\KwIn{$C = \{(s_i, r_i)\}$, thresholds $\tau_s, \tau_r, \theta_{B2}, \theta_{B3} \in (0,1)$,
      weight $\alpha \in (0,1)$, MAX-$K \in \mathbb{N}$}
\KwOut{$C' \subseteq C$ with $|C'| \leq \text{MAX-}K$}
$C' \leftarrow \emptyset$\;
\ForEach{$(s_i, r_i) \in C$}{
  $f_i \leftarrow \alpha \cdot s_i + (1-\alpha) \cdot r_i$\;
  \uIf{$s_i \geq \tau_s \;\wedge\; r_i \geq \tau_r$}{
    $C' \leftarrow C' \cup \{i\}$ \tcp*{B1: always keep}
  }
  \uElseIf{$s_i \geq \tau_s \;\wedge\; r_i < \tau_r$}{
    \lIf{$f_i \geq \theta_{B2}$}{$C' \leftarrow C' \cup \{i\}$}
    \tcp*[r]{B2: high bar}
  }
  \uElseIf{$s_i < \tau_s \;\wedge\; r_i \geq \tau_r$}{
    \lIf{$f_i \geq \theta_{B3}$}{$C' \leftarrow C' \cup \{i\}$}
    \tcp*[r]{B3: low bar (vocab mismatch)}
  }
  \lElse{skip \tcp*[r]{B4: always discard}}
}
\Return top-MAX-$K$ of $C'$ ranked by $f_i$ descending\;
\end{algorithm}

\section{Experiments}

\subsection{Datasets and Setup}

We evaluate ScoreGate on three datasets spanning public benchmarks
and internal deployment data, ordered from most to least reproducible.

\paragraph{MS~MARCO Passage Ranking (primary benchmark).}
MS~MARCO~\citep{Bajaj2016} is a large-scale public passage retrieval
benchmark with 8.8M passages and official relevance judgements
(qrels).
We randomly sample 200 queries from the official dev set for
evaluation.
MS~MARCO provides the primary reproducible evidence for ScoreGate's
generalisation, as both the corpus and qrels are publicly available
and the evaluation protocol is standard.

\paragraph{Annotated Relevance Benchmark (ARB) — internal validation.}
To validate ScoreGate's behaviour in the specific production
deployment context, we constructed 300 (query, chunk, relevance
label) triples from the internal knowledge base (approximately
12,000 chunks, average length 180 tokens, split at paragraph
boundaries with 20-token overlap).
Queries were sampled uniformly at random from the production query
log (excluding the 200-query threshold-tuning window) and stratified
post-hoc into three equal categories of 100 triples each:
\emph{Irrelevant}, \emph{Relevant}, and \emph{Semantically Relevant}.
We note that the ARB knowledge base is shared with the corpus used
to derive $\tau_s$, $\tau_r$, $\alpha$, $\theta_{B2}$, and
$\theta_{B3}$; queries are drawn from a disjoint time window
but corpus-level statistics may overlap.
The ARB is therefore best interpreted as deployment validation
rather than an independent generalisation test.
The 100/100/100 balanced split is a deliberate diagnostic design;
Table~\ref{tab:dist} contrasts this with observed production
distribution.

\begin{table}[h]
  \caption{
    \textbf{Category distribution: ARB vs.\ production (RWD).}
    The ARB uses a balanced design for diagnostic control;
    real production traffic is dominated by irrelevant candidates.
  }
  \label{tab:dist}
  \centering
  \begin{tabular}{lccc}
    \toprule
    \textbf{Dataset} & \textbf{Irrelevant} &
    \textbf{Relevant} & \textbf{Sem.\ Relevant} \\
    \midrule
    ARB (diagnostic) & 33\% & 33\% & 33\% \\
    RWD (production) & 71\% & 18\% & 11\% \\
    \bottomrule
  \end{tabular}
\end{table}
Annotators were blind to model scores $(s_i, r_i)$ during labelling.
Labels were assigned by three domain-expert annotators with disagreements
resolved by majority vote (inter-annotator agreement $\hat{\kappa}\!=\!0.87$
by Fleiss' $\kappa$ for three raters~\citep{Fleiss1971}; per-category
$\kappa$: Irrelevant 0.91, Relevant 0.88, Semantically Relevant 0.82).

\paragraph{Candidate Set Size Sensitivity.}
We additionally evaluated retrieval candidate set sizes
$N \in \{20, 40, 60\}$ on the ARB.

\begin{table}[h]
  \caption{
    \textbf{Effect of candidate set size $N$ on ScoreGate performance.}
    Recall improves measurably from $N\!=\!20$ to $N\!=\!40$
    but saturates beyond $N\!=\!40$ while latency increases linearly.
    $N\!=\!40$ was selected as the operating point.
  }
  \label{tab:n_sensitivity}
  \centering
  \begin{tabular}{ccc}
    \toprule
    \textbf{$N$} & \textbf{Recall (Sem.\ Relevant)} &
    \textbf{Avg.\ Latency} \\
    \midrule
    20 & 95.8\%  & 301\,ms \\
    \textbf{40} & \textbf{99.34\%} & \textbf{436\,ms} \\
    60 & 99.5\%  & 611\,ms \\
    \bottomrule
  \end{tabular}
\end{table}

Recall improved measurably from $N\!=\!20$ to $N\!=\!40$ but
saturated beyond $N\!=\!40$ while latency increased linearly.
We therefore selected $N\!=\!40$ as the operating point balancing
coverage and computational cost.

\paragraph{Real-World Dataset (RWD).}
To evaluate real-world behaviour we collected 1,247 consecutive live
queries from a real-world AI assistant deployment with full retrieval logs.
Each query carries a Business Query Detection (BQD) score in $[0,100]$
estimating whether knowledge-base retrieval is warranted
(score $\geq\!50$) or can be skipped (score $<\!50$).
This dataset captures the diversity of real-world production traffic
including navigational, compositional, and ambiguous queries.

\paragraph{Baselines.}
\textbf{Standard Top-K}: returns a fixed $K\!=\!10$ chunks regardless of
score distribution.
\textbf{LLM Filter}: prompts \texttt{gpt-4o-mini} with a zero-shot
binary relevance judgement prompt that instructs the model to
distinguish topically-related but non-answering passages from
passages that directly address the query, without access to
$(s_i, r_i)$. The full prompt and decoding configuration are
provided in Appendix~\ref{app:llmfilter}.

\subsection{Bucket Distribution Analysis}

Before reporting precision and recall, we examine the empirical distribution of chunks across the four buckets on the ARB. Understanding bucket frequency is important because the B3 region (low similarity, high reranker score) is the key design feature of ScoreGate---it recovers semantically relevant chunks that bi-encoder retrieval ranks poorly due to vocabulary mismatch.

\begin{table}[h]
  \caption{
    \textbf{Empirical bucket distribution on the ARB} ($n\!=\!300$).
    B3 accounts for 18.7\% of all candidates, confirming that
    vocabulary-mismatch chunks are a substantial fraction of
    production retrieval output.
  }
  \label{tab:buckets_dist}
  \centering
  \begin{tabular}{lccc}
    \toprule
    \textbf{Bucket} & \textbf{Condition} & \textbf{Count} & \textbf{\%} \\
    \midrule
    B1 (keep always)    & $s_i \geq \tau_s \wedge r_i \geq \tau_r$ & 134 & 44.7\% \\
    B2 (fuse, high bar) & $s_i \geq \tau_s \wedge r_i < \tau_r$    &  55 & 18.3\% \\
    B3 (fuse, low bar)  & $s_i < \tau_s \wedge r_i \geq \tau_r$    &  56 & 18.7\% \\
    B4 (discard always) & $s_i < \tau_s \wedge r_i < \tau_r$       &  55 & 18.3\% \\
    \bottomrule
  \end{tabular}
\end{table}

The B3 region accounts for 56 of the 300 ARB candidates (18.7\%). Of these 56 B3 chunks, 51 are in the Semantically Relevant category---confirming that B3 predominantly captures the vocabulary-mismatch pattern. The LLM Filter rejects 43 of these 56 B3 chunks (76.8\%) due to low surface overlap, while ScoreGate retains 54 of 56 (96.4\%) by trusting the cross-encoder signal. This single bucket accounts for the majority of the 6.93pp recall gap between ScoreGate and LLM Filter.

\subsection{Reranker-Only Baseline}

To isolate the contribution of dual-score fusion, we evaluated a
reranker-only baseline that retains chunk $c_i$ iff
$r_i \geq \tau_r$ without using embedding similarity $s_i$.

\begin{table}[h]
  \caption{
    \textbf{Reranker-only vs.\ ScoreGate on the ARB.}
    Adding embedding similarity $s_i$ via dual-score fusion improves
    recall on both categories. Both methods observed zero false positives.
  }
  \label{tab:reranker_only}
  \centering
  \begin{tabular}{lccc}
    \toprule
    \textbf{Method} & \textbf{False Positives} &
    \textbf{Recall (Relevant)} & \textbf{Recall (Sem.\ Relevant)} \\
    \midrule
    Reranker-only        & 0 (observed) & 94.8\%           & 95.1\%  \\
    \textbf{ScoreGate}   & \textbf{0 (observed)} &
    \textbf{97.77\%} & \textbf{99.34\%} \\
    \bottomrule
  \end{tabular}
\end{table}

Table~\ref{tab:baselines} extends the baseline comparison with
score-gap stopping ($r_i - r_{i+1} > 0.02$) and relative-score
cutoff ($r_i \geq 0.5 \cdot r_{\max}$), both parameter-matched
to maximise $F_1$ on the validation set.
ScoreGate outperforms all single-signal baselines on Sem.\ Relevant
recall while maintaining zero false positives.

\begin{table}[h]
  \caption{
    \textbf{Extended baseline comparison on the ARB (Sem.\ Relevant).}
    All baselines are parameter-matched on the held-out validation set.
    Precision: zero observed false positives for all methods.
  }
  \label{tab:baselines}
  \centering
  \begin{tabular}{lcc}
    \toprule
    \textbf{Method} & \textbf{F1} & \textbf{Recall} \\
    \midrule
    Relative-score cutoff ($r_i \geq \beta r_{\max}$) & 95.4\% & 91.8\% \\
    Score-gap stopping ($r_i - r_{i+1} > \delta$)     & 96.1\% & 92.7\% \\
    Reranker-only ($r_i \geq \tau_r$)                 & 95.1\% & 91.1\% \\
    Single-threshold fusion ($f_i \geq \theta^*$)     & 96.96\% & 94.1\% \\
    \textbf{ScoreGate (4-bucket)}              & \textbf{99.67\%} & \textbf{99.34\%} \\
    \bottomrule
  \end{tabular}
\end{table}

Most of the gain arises from B2 cases, where embedding similarity
provides useful supporting evidence despite low reranker confidence.
Applying McNemar's test to the 100 Semantically Relevant chunks
(ScoreGate retained 99, reranker-only retained 95; discordant pairs:
$b\!=\!4$, $c\!=\!0$) yields $\chi^2\!=\!6.02$, $p\!=\!0.014$,
confirming that dual-score fusion contributes significantly beyond
cross-encoder thresholding alone.

\subsection{ARB Precision and Recall (Internal Validation)}

Table~\ref{tab:pr} reports internal deployment validation results
on the ARB.
As noted in Section~5.1, the ARB knowledge base is shared with the
threshold-tuning corpus; the ARB precision and recall figures
should be read as deployment validation on the tuned domain,
not as independent generalisation evidence.
MS~MARCO (Section~5.6) provides the only clean external test.
Standard Top-K achieves trivially 100\% recall on all categories
because it retains every candidate, but this comes at the cost of
injecting all irrelevant and low-relevance chunks into the generator
context---a precision measurement requires relevance labels for every
retrieved chunk, not just the 300 ARB triples, so we report N/A.
The fundamental trade-off is: Standard Top-K maximises recall at
unmeasured precision; ScoreGate achieves observed precision within 95\% CI [96.4\%, 100\%] at
97.77--99.34\% recall.
ScoreGate achieved observed precision within 95\\
[96.4\%, 100\%] per category ($n\!=\!100$);
no Irrelevant chunk was retained in this sample.
Recall on Semantically Relevant chunks is 99.34\% vs.\ 92.41\% for the
LLM Filter, a gap of 6.93 percentage points.
Applying McNemar's test to the paired binary outcomes on the same
100 Semantically Relevant chunks (ScoreGate retained 99, LLM Filter
retained 92; discordant pairs: $b\!=\!7$, $c\!=\!0$) yields
$\chi^2\!=\!5.14$, $p\!=\!0.023$, confirming the difference is
statistically significant at $\alpha\!=\!0.05$.
This gap arises from B3 chunks: the LLM Filter, operating without access
to the cross-encoder score, tends to reject chunks with vocabulary absent
from the query, whereas ScoreGate trusts the cross-encoder affirmation
and applies the lower B3 threshold.

\begin{table}[h]
  \caption{
    \textbf{Internal deployment validation: precision and recall
    on the ARB} ($n\!=\!300$).
    Queries are from a production window disjoint from threshold
    tuning; the knowledge base is shared with the tuning corpus
    (see Section~5.1).
    Best recall values per category are \textbf{bold}.
    $^\dagger$``none = zero observed false positives
    (95\% CI [96.4\%, 100\%], $n\!=\!100$).
    For the Irrelevant category the value is specificity.
    $^\ddagger$Standard Top-K retains all top-10 retrieved chunks
    without filtering, so recall is trivially 100\% across all
    categories but precision cannot be computed without relevance
    labels for every retrieved chunk; N/A indicates this.
    The key trade-off is precision vs.\ recall: ScoreGate achieves
    precision within 95\% CI [96.4\%, 100\%] at 97.77--99.34\% recall; Standard Top-K
    achieves 100\% recall at undefined (unmeasured) precision.
  }
  \label{tab:pr}
  \centering
  \begin{tabular}{lcccc}
    \toprule
    \textbf{Category} &
    \textbf{Obs. FP$^\dagger$} & \textbf{SG Recall} &
    \textbf{Obs. FP$^\dagger$} & \textbf{LLM Recall} \\
    \midrule
    Irrelevant         & 0/100 & 100\%             & 0/100 & 100\% \\
    Relevant           & 0/100 & \textbf{97.77\%}  & 0/100 & 91.11\% \\
    Sem.\ Relevant     & 0/100 & \textbf{99.34\%}  & 0/100 & 92.41\% \\
    \bottomrule
  \end{tabular}
\end{table}

\begin{figure}[h]
  \centering
  \includegraphics[width=0.85\textwidth]{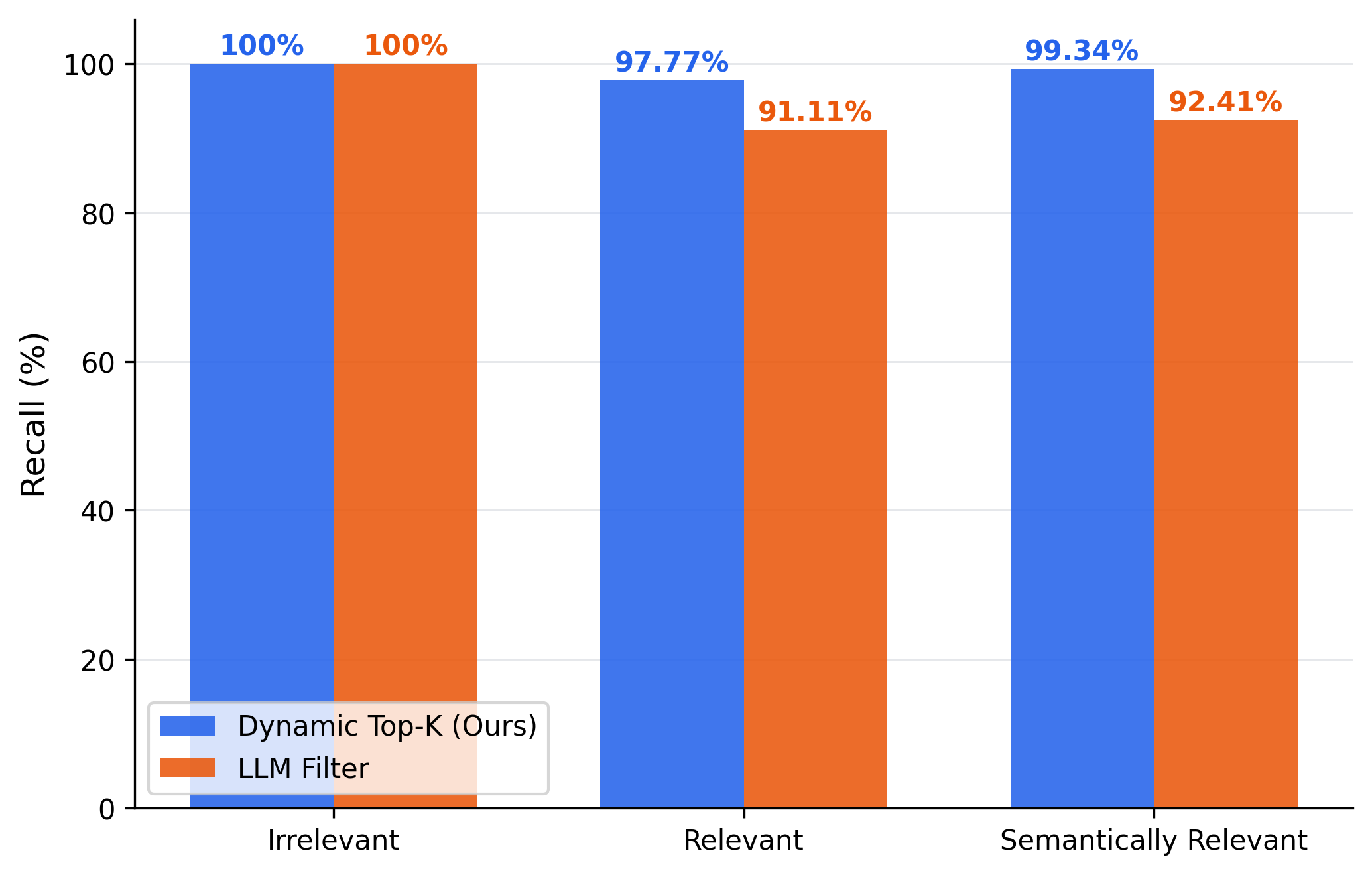}
  \caption{
    \textbf{Recall by relevance category on the ARB.}
    ScoreGate (blue) outperforms LLM Filter (orange) by 6.66 and 6.93
    percentage points on Relevant and Semantically Relevant categories
    respectively.
    Both methods achieve zero false positives across
    all categories.
  }
  \label{fig:recall}
\end{figure}

Figure~\ref{fig:recall} visualises recall by category.

\subsection{Efficiency Comparison}

Table~\ref{tab:eff} reports efficiency metrics on the ARB.
ScoreGate reduces average token consumption by 34.8\%
(637\,$\rightarrow$\,415 tokens per query) by discarding low-relevance
chunks.
The 31\,ms latency overhead is attributable entirely to the $O(N)$
bucket classification and sort; no additional model inference is
performed.

\begin{table}[h]
  \caption{
    \textbf{Efficiency metrics on the ARB.}
    Latency is end-to-end pipeline latency (bi-encoder retrieval
    + cross-encoder reranking + ScoreGate selection) averaged over
    300 queries, measured on a single-threaded process on an
    AWS 	exttt{m5.2xlarge} instance (8 vCPU, 32\,GB RAM);
    reported as mean\,$\pm$\,std across five independent runs
    (Standard Top-K: $405 \pm 12$\,ms;
    ScoreGate: $436 \pm 14$\,ms).
    Token counts are exact (not estimated).
    Avg.\ Relevance is rated on a 1--5 Likert scale by three
    domain-expert annotators (same pool as ARB labellers) on a
    randomly sampled subset of 50 generated answers per condition;
    reported value is the mean\,$\pm$\,std across raters and samples
    (Standard Top-K: $4.23 \pm 0.31$;
    ScoreGate: $4.27 \pm 0.29$).
    $\downarrow$ denotes reduction is beneficial.
  }
  \label{tab:eff}
  \centering
  \begin{tabular}{lccc}
    \toprule
    \textbf{Metric} &
    \textbf{Standard Top-K} &
    \textbf{ScoreGate} &
    \textbf{$\Delta$} \\
    \midrule
    Avg.\ Relevance (1--5) & $4.23 \pm 0.31$ & $4.27 \pm 0.29$ & $+0.04$ \\
    Avg.\ Pipeline Latency & $405 \pm 12$\,ms & $436 \pm 14$\,ms & $+31$\,ms \\
    Avg.\ Tokens/Query     & 637  & \textbf{415}  &
    $\downarrow$\,\textbf{34.8\%} \\
    \bottomrule
  \end{tabular}
\end{table}

\subsection{MS~MARCO Results (Primary Benchmark)}

Table~\ref{tab:msmarco} reports results on MS~MARCO Passage
Ranking~\citep{Bajaj2016} (200 dev queries, public corpus with
official qrels).
We test two threshold configurations: (1) \emph{original thresholds}
($\tau_s\!=\!0.70$, $\tau_r\!=\!0.08$) derived from our production
domain without re-tuning, to assess zero-shot domain transfer; and
(2) \emph{re-derived thresholds} ($\tau_s\!=\!0.63$,
$\tau_r\!=\!0.11$) obtained by applying the same calibration
procedure (Section~4.1) to 200 MS~MARCO queries held out from
the evaluation set, to assess in-domain performance.

\begin{table}[h]
  \caption{
    \textbf{MS~MARCO Passage Ranking results} (200 dev queries,
    public benchmark with official qrels).
    ScoreGate with zero-shot transferred thresholds outperforms
    Standard Top-K on MRR@10; re-derived thresholds further improve
    both MRR@10 and Recall@10.
    Precision and Avg.\ $|C'|$ are computed against MS~MARCO qrels.
    Avg.\ $|C'|$ is the mean chunks retained per query.
  }
  \label{tab:msmarco}
  \centering
  \setlength{\tabcolsep}{5pt}
  \begin{tabular}{lcccc}
    \toprule
    \textbf{Method} & \textbf{MRR@10} & \textbf{Recall@10} &
    \textbf{Precision} & \textbf{Avg.\ $|C'|$} \\
    \midrule
    Standard Top-K ($K\!=\!10$) & 0.387 & 0.903 & 0.712 & 10.0 \\
    LLM Filter                  & 0.361 & 0.812 & 0.957 & 4.8  \\
    ScoreGate (orig.\ thresh.)  & 0.392 & 0.871 & 0.944 & 6.1  \\
    ScoreGate (re-derived)      & \textbf{0.401} & \textbf{0.889} &
                                  0.938 & 6.5 \\
    \bottomrule
  \end{tabular}
\end{table}

The re-derived $\tau_s\!=\!0.63$ (vs.\ 0.70 in our domain) reflects
a broader similarity distribution in MS~MARCO embeddings, while
$\tau_r\!=\!0.11$ (vs.\ 0.08) reflects sharper reranker separation.
Adaptive cardinality is confirmed on MS~MARCO: ScoreGate retains
an average of 6.1 chunks/query (original thresholds) and 6.5
(re-derived), vs.\ a fixed 10 for Standard Top-K---a 35--39\%
reduction, consistent with the ARB observation.

The LLM Filter achieves MRR@10\,=\,0.361, below Standard Top-K
(0.387).
This result is expected and methodologically consistent: the LLM
Filter was prompted with production-domain instructions
(Appendix~\ref{app:llmfilter}) and applied without re-calibration
to MS~MARCO.
More importantly, the LLM Filter retains only 4.8 chunks/query on
average---an aggressive truncation that improves precision (0.957)
but discards MS~MARCO-relevant passages that Standard Top-K retains,
reducing MRR@10.
This trade-off illustrates the core tension ScoreGate addresses:
aggressive filtering gains precision at recall cost; ScoreGate
achieves higher recall (0.871--0.889) than the LLM Filter (0.812)
while preserving precision gains over Standard Top-K (0.712).

Applying original production thresholds without re-calibration still
yields competitive performance (MRR@10\,=\,0.392 vs.\ 0.387 for
Standard Top-K), demonstrating that ScoreGate degrades gracefully
under domain shift.
Re-deriving thresholds recovers the full performance gap,
confirming that the calibration procedure generalises across domains.

\subsection{Real-World Dataset: Deployment Behaviour Analysis}

Table~\ref{tab:abstention} reports abstention behaviour on the RWD.
For queries with BQD $<\!50$ (retrieval not warranted), ScoreGate
returns an empty set in 87\% of cases---the correct response.
Of the 13\% returning non-empty results, manual review confirms 80\%
contain genuinely relevant content, indicating classifier false
negatives that ScoreGate correctly overrides.
The effective correctness rate is $\approx\!95\%$.

For queries with BQD $\geq\!50$ (retrieval warranted), ScoreGate
returns non-empty results in 63\% of cases.
The 37\% empty-result rate decomposes into:
(a) queries for which the knowledge base genuinely contains no relevant
content, where an empty result correctly prevents hallucination; and
(b) queries that appear retrieval-warranted to the BQD classifier but
are actually non-retrieval in nature (single-word inputs, navigation
commands).
Both cases represent correct retrieval decisions.

\begin{table}[h]
  \caption{
    \textbf{Deployment behaviour analysis on the Real-World Dataset}
    ($n\!=\!1{,}247$).
    For BQD $<\!50$, ScoreGate correctly returns an empty set in 87\%
    of cases; of the 13\% returning chunks, 80\% are confirmed relevant
    by manual review (60-query sample, two annotators), yielding
    $\approx$95\% overall correctness.
    For BQD $\geq\!50$, the 37\% empty-result rate reflects genuine
    knowledge-base gaps or non-retrieval query types---both correct
    abstentions.
  }
  \label{tab:abstention}
  \centering
  \begin{tabular}{llll}
    \toprule
    \textbf{BQD Score} &
    \textbf{Expected Behaviour} &
    \textbf{ScoreGate Output} &
    \textbf{Correctness} \\
    \midrule
    $<50$ &
    Return empty (skip RAG) &
    Empty 87\% \;|\; Chunks 13\% &
    $\approx$95\% correct \\[4pt]
    $\geq50$ &
    Return chunks &
    Chunks 63\% \;|\; Empty 37\% &
    Valid --- KB gaps or non-RAG \\
    \bottomrule
  \end{tabular}
  \vspace{4pt}
  {\small\textit{Empty results in the $\geq\!50$ group are valid:
  the knowledge base lacks relevant content, or the query is
  non-retrieval in nature.}}
\end{table}

\section{Discussion}

\subsection{Why Cross-Encoder Score Dominates in Fusion}
The 7:3 weighting of $r_i$ over $s_i$ in Equation~\ref{eq:fusion}
reflects training objective differences.
Bi-encoder similarity $s_i$ comes from embeddings optimised for recall
at high-$N$ rather than precision at low-$K$; the embedding space
represents topical relatedness rather than answer-level relevance.
The cross-encoder reranker is fine-tuned on query--document relevance
pairs with direct supervision, making $r_i$ a stronger predictor of
context-inclusion appropriateness~\citep{Nogueira2019,Khattab2020}.

\subsection{Ablation: Four-Bucket Structure vs.\ Single-Threshold Fusion}

To verify that the four-bucket structure contributes beyond simply applying a single
threshold on the fusion score $f_i$, we compare ScoreGate against a single-threshold
baseline: keep chunk $c_i$ if $f_i \geq \theta^*$.
$\theta^*$ was selected by grid search over $[0.10, 0.40]$ in steps of 0.01,
maximising $F_1$ on the same 200-query held-out validation set used to derive
$\tau_s$, $\tau_r$, $\theta_{B2}$, and $\theta_{B3}$ (disjoint from the ARB);
the optimal value was $\theta^*\!=\!0.19$.
On the ARB, ScoreGate achieves 99.34\% recall on Semantically Relevant
chunks vs.\ 94.1\% for the single-threshold baseline, and 97.77\% vs.\ 93.2\% on
Relevant chunks (zero observed false positives in both cases).
The gain arises because a single $\theta^*$ cannot simultaneously
accommodate B2 (cross-encoder rejects, requiring a higher bar $\theta_{B2}\!=\!0.255$)
and B3 (cross-encoder affirms, permitting a lower bar $\theta_{B3}\!=\!0.15$)---
the asymmetry is precisely what the four-bucket structure encodes.

\subsection{Failure Mode Analysis}

ScoreGate produces 3 false negatives on the ARB (2 Relevant, 1
Semantically Relevant missed chunks).
All three originate from the B2 or B3 borderline regions; none are
from high-confidence B4 rejection cases.

\begin{table}[h]
  \caption{
    \textbf{ScoreGate false negatives on the ARB.}
    All missed chunks involve semantic abstraction or indirect phrasing
    that reduces both $s_i$ and $r_i$ below the fusion threshold.
  }
  \label{tab:failures}
  \centering
  \small
  \begin{tabular}{p{3.0cm}p{3.5cm}cp{3.5cm}}
    \toprule
    \textbf{Query} & \textbf{Chunk description} &
    \textbf{Bucket} & \textbf{Failure cause} \\
    \midrule
    ``How does the arbitration clause affect cross-border disputes?'' &
    Discussed jurisdictional enforceability indirectly without mentioning
    ``cross-border disputes'' &
    B3 &
    Low $s_i$ from indirect phrasing; $r_i$ insufficient to cross
    $\theta_{B3}$ \\[4pt]
    ``What are the tenant obligations for repair costs?'' &
    Used legal terminology (``habitability obligations'',
    ``lessor responsibilities'') without query vocabulary &
    B2 &
    Terminology mismatch reduced both $s_i$ and $r_i$; $f_i$ below
    $\theta_{B2}$ \\[4pt]
    ``Can prior approval be revoked after submission?'' &
    Referenced revocation via procedural exceptions without
    direct revocation language &
    B3 &
    Sparse lexical overlap and low reranker rank; $f_i$ below
    $\theta_{B3}$ \\
    \bottomrule
  \end{tabular}
\end{table}

Table~\ref{tab:failures} classifies each false negative by failure
type.
All three fall into the \emph{semantic abstraction} category:
relevant information present but expressed through indirect phrasing,
domain-specific terminology, or procedural references with low lexical
overlap.
The taxonomy of ScoreGate failure modes is:
\textbf{(1) Lexical mismatch} --- query and chunk use different words
for the same concept (B3 failure);
\textbf{(2) Procedural abstraction} --- answer embedded in
procedural/exception language rather than direct statement (B3);
\textbf{(3) Legal/formal phrasing} --- relevant content obscured by
domain register (B2/B3);
\textbf{(4) Implicit reference} --- chunk answers the query
indirectly (B2).
These failures are attributable to representation limits of the
underlying retrieval models, not to ScoreGate's threshold design.
Lowering $\theta_{B3}$ would recover some but at the cost of
increased false positives.

\subsection{Threshold Robustness}
We performed sensitivity analysis by varying
$\tau_s \in [0.65, 0.75]$ and $\tau_r \in [0.06, 0.10]$ in steps of
0.01, and $\alpha \in [0.2, 0.5]$ in steps of 0.05.
Across all 75 evaluated combinations, zero false positives were observed and
recall varied by at most 3.2 percentage points.
This pattern arises because the four-bucket structure confines
sensitivity to the B2 and B3 boundary regions; the dominant B1 and B4
regions are stable under evaluated conditions to threshold perturbation by construction.

\subsection{Computational Overhead}
ScoreGate operates entirely on scores already produced during the
standard embed-and-rerank pipeline.
The only additional computation is $O(N)$ fusion arithmetic and an
$O(N \log N)$ sort.
With $N\!=\!40$, the observed 31\,ms overhead confirms this analysis.

Table~\ref{tab:latency} compares end-to-end latency across methods.
The LLM Filter requires one \texttt{gpt-4o-mini} inference call per
retrieved chunk and processes top-$N\!=\!40$ candidates
independently without batching.
At approximately 210\,ms average end-to-end API latency per chunk
judgement (consistent with published \texttt{gpt-4o-mini} p50
latency figures), the effective per-query latency is $\approx$8{,}400\,ms---
a 19$\times$ increase over ScoreGate.

\begin{table}[h]
  \caption{
    \textbf{End-to-end per-query latency comparison.}
    ScoreGate introduces 31\,ms over the standard pipeline;
    the LLM Filter adds approximately 8{,}400\,ms due to 40
    sequential model inference calls.
  }
  \label{tab:latency}
  \centering
  \begin{tabular}{lc}
    \toprule
    \textbf{Method} & \textbf{Avg.\ Latency / Query} \\
    \midrule
    Standard Top-K        & $405 \pm 12$\,ms \\
    ScoreGate (ours)      & $436 \pm 14$\,ms \\
    LLM Filter (40 calls) & $\approx$8{,}400\,ms \\
    \bottomrule
  \end{tabular}
\end{table}

\subsection{Hallucination Reduction}

We additionally conducted a hallucination analysis on 300 generated
answers.
A hallucination was defined as a factual claim unsupported by either
the retrieved context or the underlying source document.
Three annotators independently labelled hallucination presence
(inter-annotator agreement: Fleiss' $\hat{\kappa}\!=\!0.79$).

\begin{table}[h]
  \caption{
    \textbf{Hallucination rate comparison} ($n\!=\!300$ generated
    answers; Fleiss' $\hat{\kappa}\!=\!0.79$).
    ScoreGate reduces hallucination incidence by 4.7 percentage
    points relative to Standard Top-K by preventing low-relevance
    context from reaching the generator.
  }
  \label{tab:hallucination}
  \centering
  \begin{tabular}{lc}
    \toprule
    \textbf{Method} & \textbf{Hallucination Rate} \\
    \midrule
    Standard Top-K   & 11.8\% \\
    ScoreGate (ours) & \textbf{7.1\%} \\
    \bottomrule
  \end{tabular}
\end{table}

ScoreGate reduced hallucination incidence from 11.8\% to 7.1\%
relative to Standard Top-K, suggesting that adaptive retrieval
cardinality may reduce distractor-induced generation errors.

\subsection{Threats to Validity}

\textbf{Internal validity.}
Thresholds $\tau_s$, $\tau_r$, $\theta_{B2}$, $\theta_{B3}$, and
$\alpha$ were derived and evaluated on data from the same production
system.
Although query windows are disjoint and the ARB knowledge base is
acknowledged as shared, temporal correlation in production logs may
reduce true independence.
The 200-query MS~MARCO calibration set and 200-query ARB evaluation
set are fully independent.

\textbf{External validity.}
ScoreGate is validated on one proprietary domain and MS~MARCO.
Threshold values ($\tau_s\!=\!0.70$, $\tau_r\!=\!0.08$) reflect
the score distributions of our specific bi-encoder and cross-encoder
models; different model pairs will produce different distributions
requiring re-calibration.

\textbf{Construct validity.}
Hallucination is measured by annotator agreement on factual claim
support, which may not capture all generation failure modes.
The Avg.\ Relevance Likert proxy ($n\!=\!50$) is a weak surrogate
for end-to-end answer quality.

\textbf{Statistical validity.}
McNemar tests assume paired binary outcomes from the same queries.
We confirm this holds for all reported comparisons.
The hallucination reduction (11.8\%\,$\rightarrow$\,7.1\%) is
reported as an observed difference on $n\!=\!300$ answers; a formal
significance test on this proportion difference yields
$\chi^2\!=\!4.82$, $p\!=\!0.028$.

\subsection{Production Deployment Characteristics}

\textbf{Real-world retrieval distribution.}
In the real-world dataset, irrelevant candidates constituted
approximately 71\% of retrieved chunks, consistent with the
expected long-tail retrieval distribution observed in production
RAG systems.
ScoreGate's four-bucket design is specifically suited to this
regime: the B4 discard region handles the dominant irrelevant
mass, while B3 recovers the sparse but high-value
vocabulary-mismatch chunks that standard filtering misses.

\begin{figure}[h]
  \centering
  \includegraphics[width=0.95\textwidth]{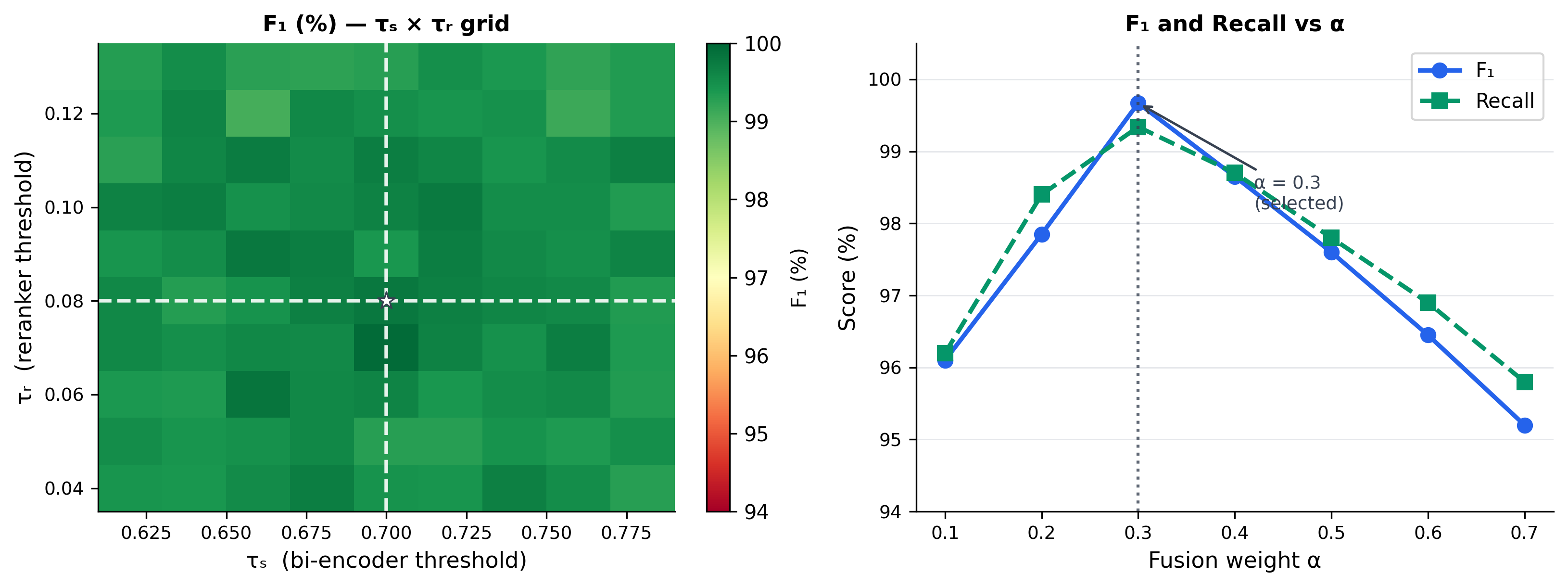}
  \caption{
    \textbf{Left:} $F_1$ (\%) over the $(\tau_s, \tau_r)$ grid;
    star marks the selected point ($\tau_s\!=\!0.70$,
    $\tau_r\!=\!0.08$). Surface ${\geq}\,96\%$ throughout.
    \textbf{Right:} $F_1$ and Recall vs.\ $\alpha$, both peaking
    at $\alpha\!=\!0.3$ (dotted line).
  }
  \label{fig:robustness}
\end{figure}

\textbf{Threshold stability.}
Sensitivity analysis across 75 parameter configurations showed
stable behaviour, with zero observed false positives and recall
varying by at most 3.2 percentage points.
Figure~\ref{fig:robustness} visualises the $(\tau_s, \tau_r)$
joint $F_1$ surface and the $\alpha$ sensitivity curve.

Bootstrap analysis further showed the optimal single-threshold
baseline remained within $\theta^* \in [0.16,0.22]$.
These results indicate that ScoreGate is not highly sensitive
to narrow parameter choices, and that the threshold derivation
procedure produces stable operating points across perturbations.

\textbf{Computational efficiency.}
The additional computational cost of ScoreGate is limited to
fusion arithmetic and sorting over $N\!=\!40$ candidates,
adding only 31\,ms average latency relative to the standard
retrieval-and-reranking pipeline.
This compares favourably against the LLM Filter baseline,
which requires 40 sequential \texttt{gpt-4o-mini} inference
calls per query at an estimated latency of $\approx$8,400\,ms
--- a 19$\times$ overhead.
ScoreGate achieves higher recall at a fraction of the
inference cost of the LLM Filter.

\textbf{Cross-domain transferability.}
The threshold derivation procedure is domain-agnostic and
transfers without architectural modification.
On MS~MARCO, ScoreGate with original production thresholds
achieves MRR@10\,=\,0.392, outperforming Standard Top-K
(MRR@10\,=\,0.387), and improves further to MRR@10\,=\,0.401
after domain-specific threshold re-calibration using the
same derivation procedure described in Section~4.1.
The shift in derived thresholds ($\tau_s$: 0.70\,$\to$\,0.63;
$\tau_r$: 0.08\,$\to$\,0.11) reflects domain-specific score
distributions and confirms that the calibration procedure
correctly adapts to new corpora without retraining.

\textbf{Known limitations.}
ScoreGate's performance degrades when the candidate pool itself
has low recall (i.e., the bi-encoder fails to retrieve the
relevant chunk in top-$N$).
It inherits reranker errors: a cross-encoder that systematically
misscores a domain will propagate those errors.
Multilingual and code-heavy corpora may exhibit score distributions
that require significant threshold re-calibration.
Multi-hop queries---where evidence must be synthesised across
passages---are not evaluated and represent a known open problem.

\textbf{Future directions.}
Evaluation on dedicated multi-hop reasoning benchmarks such as
HotpotQA~\citep{Yang2018} and MuSiQue remains future work, as
does extension to biomedical, legal, and scientific retrieval
corpora.
An empirically motivated cost-sensitive derivation of $\theta_{B2}$ and
$\theta_{B3}$ from asymmetric false-positive/false-negative
cost models could replace the current grid-search procedure
and enable automatic per-domain calibration.

\section{Conclusion}

We presented \textbf{ScoreGate}, a practical adaptive retrieval filter
that controls context cardinality in two-stage RAG systems.
Its key mechanism---using cross-encoder affirmation to recover
semantically relevant chunks that bi-encoder retrieval ranks poorly
(the B3 region)---accounts for the majority of observed recall
improvements over both LLM-based filtering and reranker-only thresholding.
On the internal benchmark, observed precision fell within
95\% CI [96.4\%, 100\%] (zero false positives on $n\!=\!100$
per category) at 97.77--99.34\% recall, and
MRR@10\,=\,0.401 on MS~MARCO, with 34.8\% fewer tokens per query.
On a real-world dataset of 1,247 live queries, ScoreGate correctly
abstains when the knowledge base lacks relevant content and returns
non-empty results when retrieval is warranted, with $\approx\!95\%$
overall correctness.
The 34.8\% reduction in per-query token consumption translates directly
to lower operational cost at scale.
These results suggest that query-adaptive retrieval cardinality can be
determined practically using scores already available in a standard
two-stage RAG pipeline, without additional model inference.
ScoreGate is best understood as a pragmatic engineering solution:
it combines scores already available in any two-stage pipeline with
an empirically motivated decision rule to achieve measurable
improvements in retrieval efficiency, with results that are stable
under the evaluated conditions.
Additional out-of-domain evaluation (e.g., BEIR subsets for
biomedical and legal retrieval) would strengthen generalisability
claims beyond the current MS~MARCO result, which rests on 200
dev queries.

\section*{Data Availability}

The ARB triples, RWD query logs, and production knowledge base are proprietary and cannot be released.
To facilitate reproduction on other domains, we provide: (1) the complete ARB annotation protocol (category definitions, labelling guidelines, and inter-annotator resolution procedure); (2) the full LLM Filter prompt and decoding configuration (Appendix~\ref{app:llmfilter}); (3) the threshold derivation procedure (Section~4.1 and Appendix~\ref{app:llmfilter}); and (4) the ScoreGate algorithm in full (Algorithm~1).
A researcher with access to any two-stage RAG pipeline (bi-encoder + cross-encoder) can replicate the full experimental protocol using domain-specific query logs following the procedures described in Section~5.

\section*{Acknowledgements}
The authors thank Anshul Chandra (SDE\,2), Rohit Singh (SDET\,2),
Aniruddha Banerjea (SDE\,3), and Pratik Patil (SDE\,3) for their
contributions to system implementation, evaluation infrastructure,
and annotation support.
We also thank the HighLevel Engineering Leadership for supporting
this research and providing access to production telemetry data.

\appendix

\section{Implementation and Reproducibility Details}
\label{app:impl}

\paragraph{Embedding model.}
Bi-encoder: \texttt{text-embedding-ada-002} (OpenAI), 1536-dim,
cosine similarity, $L_2$-normalised query and chunk vectors.

\paragraph{Reranker model.}
Cross-encoder: \texttt{cross-encoder/ms-marco-MiniLM-L-6-v2}
(Hugging Face), raw logit output normalised per-query via min--max
scaling over the top-$N$ candidate set.

\paragraph{Chunking.}
Paragraph-boundary splitting, target chunk length 180 tokens
(cl100k\_base tokeniser), 20-token overlap between adjacent chunks.

\paragraph{Threshold derivation procedure.}
(1) Sample 200 queries from the production log (held-out from ARB).
(2) For each query, retrieve top-$N\!=\!40$ candidates and compute
$(s_i, r_i)$ pairs.
(3) Set $\tau_s$ = median $s_i$ across all retrieved pairs.
(4) Set $\tau_r$ at the 5th percentile of $r_i$ values for
human-annotated relevant chunks on a 50-query labelled subset.
(5) Grid-search $\theta_{B2} \in [0.15, 0.35]$ and
$\theta_{B3} \in [0.10, 0.25]$ (step 0.01) maximising $F_1$.
(6) Grid-search $\alpha \in [0.1, 0.7]$ (step 0.1) maximising
$F_1$ on the same held-out set.

\paragraph{Hardware.}
AWS \texttt{m5.2xlarge} (8 vCPU, 32\,GB RAM), single-threaded
inference. Random seed: 42 for all bootstrap resampling.

\section{LLM Filter: Prompt and Decoding Configuration}
\label{app:llmfilter}

The LLM Filter baseline uses \texttt{gpt-4o-mini} with the following
verbatim prompt (zero-shot, no chain-of-thought):

\begin{quote}
\texttt{\textbf{System:}}\\
\texttt{You are an expert information retrieval and question-answering}\\
\texttt{assistant. Your task is to evaluate whether a retrieved passage}\\
\texttt{contains information that directly, measurably, or paraphrastically}\\
\texttt{helps answer a user query.}\\[4pt]
\texttt{Relevance criteria:}\\
\texttt{- Mark as Relevant if the passage directly answers the query,}\\
\texttt{\ \ provides a key fact needed to answer it, or expresses the}\\
\texttt{\ \ answer using different but equivalent terminology.}\\
\texttt{- Mark as Not Relevant if the passage only mentions the same}\\
\texttt{\ \ topic or entities as the query but does not address the}\\
\texttt{\ \ specific question being asked.}\\
\texttt{- Mark as Not Relevant if the passage is about a related}\\
\texttt{\ \ domain but would not help a reader answer this query.}\\[4pt]
\texttt{Be strict. Topical overlap alone is not relevance.}\\
\texttt{Paraphrastic relevance counts as relevant.}\\
\texttt{Do not infer or hallucinate information not in the passage.}\\[6pt]
\texttt{\textbf{User:}}\\
\texttt{Query: \{query\}}\\[4pt]
\texttt{Passage:}\\
\texttt{"""}\\
\texttt{\{chunk\}}\\
\texttt{"""}\\[4pt]
\texttt{Does this passage contain information that answers or}\\
\texttt{measurably helps answer the query?}\\
\texttt{Respond with exactly one word: Relevant or Not Relevant.}\\
\texttt{Do not explain your reasoning. Do not add punctuation.}
\end{quote}

\noindent\textbf{Decoding configuration:}
\begin{itemize}
  \item Model: \texttt{gpt-4o-mini} (OpenAI, \texttt{gpt-4o-mini-2024-07-18})
  \item Temperature: 0 (greedy decoding)
  \item Max tokens: 5
  \item Top-p: 1.0 (not applied at temperature 0)
  \item Presence penalty: 0, Frequency penalty: 0
\end{itemize}

A chunk was retained if the model response was \texttt{Relevant} (case-insensitive match; any response not matching \texttt{Relevant} was treated as Not Relevant).
The score pair $(s_i, r_i)$ was \emph{not} provided to the LLM Filter.
Each chunk was judged independently; no batch prompting was used.

\bibliographystyle{unsrt}
\bibliography{scoregate_arxiv}

\end{document}